\documentclass[aps,pre,showpacs,amsmath,amssymb]{revtex4}
\usepackage{bm}
\usepackage{graphics}

\newcommand{\pp}[1]{\phantom{#1}}
\newcommand{\be}{\begin{eqnarray}}
\newcommand{\ee}{\end{eqnarray}}
\newcommand{\ve}{\varepsilon}

\newcommand{\ba}{\begin{array}}
\newcommand{\ea}{\end{array}}
\newcommand{\intinf}{\int_{-\infty} ^{+\infty} }

\begin{document}

\title{
The fundamental problem of quantum cryptography
}
\author{Marcin Paw\l owski and Marek Czachor}

\address{
Katedra Fizyki Teoretycznej i Metod Matematycznych\\
Politechnika Gda\'nska, 80-952 Gda\'nsk, Poland}

\begin{abstract}
It is known that statistical predictions of quantum theory do not depend on its interpretation. In particular, an experiment cannot distinguish between the Copenhagen interpretation (involving no hidden variables) and the de Broglie-Bohm interpretation based on nonlocal hidden variables.
Quantum cryptographic protocols, such as BB84 or E91, are secure and mutually equivalent as long as one works within the framework of Copenhagen interpretation. But they are inequivalent and insecure if one considers attacks allowed by the de Broglie-Bohm interpretation. The fundametal problem of quantum cryptography is therefore this: Are all the statements about security of quantum protocols based on our {\it belief\/} in one of the two allowed interpretations of quantum mechanics? We show that this is not the case. Ekert-type protocols can be modified in a way that makes them secure even if the de Broglie-Bohm nonlocal hidden variables exist. Bennett-Brassard-type cryptography does not seem to allow for such a correction.
\end{abstract}

\pacs{?????}

\maketitle

\section{Introduction}

Our confidence in classical public key cryptosystems such as RSA is based on a {\it belief\/} that there is no efficient algorithm that allows one to find the private key given only the public one. Since there is no proof that such a classical algorithm does not exist, and we know that quantum Shor's algorithm \cite{Shor} is capable of doing just that, we cannot consider RSA fundamentally secure. Hence the need for quantum cryptosystems which should, in principle, be safe against any attack. Protocols such as BB84 and E91 are believed to fulfill the criterion. The first of them takes its strength from the von Neumann postulate, since gaining any knowledge about a system being sent would result in wavefunction collapse and loss of information carried by the system \cite{BB}, and this is detectable. Ekert's protocol bases on nonlocal correlations within the system \cite{E91}. Here any attempt to extract information changes the initial entangled state decreasing the Bell expectation value. Both protocols assume that the Copenhagen interpretation is true. Such an assumption is not fully justified since there are other interpretations of quantum mechanics that produce identical observable results. 

In the de Broglie-Bohm approach (which so far predicts outcomes undistinguishable form those given by the Copenhagen interpretation \cite{2-time}) there is no need to invoke the wavefunction collapse hypothesis. A hidden-variable description of Stern-Gerlach measurements for both a single spin and a pair of spins in a singlet state is given in detail in the monograph \cite{Holland}. The results of measurements can be deduced from the position of the particle trajectory (a hidden variable). An eavesdropper knows for sure and in advance the results of spin measurements. Accordingly, the usual protocols are insecure.
Apparently, the situation is not dramatically better than in the case of classical cryptography; the fact that we may not believe in nonlocal hidden variables is not yet a guarantee of fundamental security of our quantum cryptosystems. 

In the present paper we show that this is not the end of the story. 
We prove that even in an interpretation that includes nonlocal hidden variables it is possible to create a secure quantum key distribution protocol. The crucial idea is to take advantage of the possibility of flipping-over the spin of particle in the lab of Alice, say, by a measurement performed by Bob in his lab. The effect is nonlocal, preserves one-to-one correlations typical of the singlet state, and is possible if and only if the two particles are in an entangled state. In such a protocol the knowledge of particle positions in transit between the source and the labs of Alice and Bob is not sufficient for determining the key. The key is literally created at the very last moment when the particles are already in the labs of Alice and Bob. The security is here a consequence of nonlocality. 

The fact that such a possibility may exist followed from the earlier study of nonlocal hidden-variables models of spin-1/2 and the singlet state \cite{MC}. Still, the toy models of EPR correlations were not reliable enough. We needed a hidden-variable model that is a candidate for a theory equivalent to standard quantum mechanics. This is why in the present paper we have concentrated on the de Broglie-Bohm model. The analysis follows, with slight modifications, the one given in \cite{Holland}.

\section{Double Stern-Gerlach experiment}

We consider an experiment where a pair of particles in the singlet state is sent along the 
$y$ axis to Alice and Bob. Each of the particles enters then the Stern-Gerlach apparatus. The magnetic fields in both devices are aligned in the same direction, and each particle spends in the field the same time $T$. Both particles have mass $m$ and magnetic moment $\mu$. We make the following standard assumptions:
\begin{itemize}
\item We suppose that the magnetic fields are large enough to neglect the kinetic part of the Hamiltonian during the time of interaction of a particle with the fields.
\item We also suppose that the time $T$ is short enough not to change particles position in the $y$-direction. During the interaction with the field the particle is only given velocity in $z$-direction.
\item We will consider fields of the form: $B(x,y,z)=B(z)=B_0+Bz$. Such fields are clearly non-physical, but their use is justified by B\"{o}hm in \cite{Bohm79}.
\item We assume that particles' motion along the $y$ axis is constant throughout the experiment.
\end{itemize}
Both assumptions and calculations below are based on those given by Holland in \cite{Holland}.

We take the fields in Alice's device to be $B_1(z)$ and in Bob's $B_2(z)$, and given explicitly by
\be
B_1(z)&=& B_0+Bz \\
B_2(z)&=& sK(B_0+Bz)
\ee
where we assume $B>0$, $K>1$ and $s=\pm 1$ is a number randomly chosen by Bob.

The initial wavefunction at the entrance is in the singlet state
\be \label{1}
\psi_0(z_1,z_2)=f_1(z_1)f_2(z_2)\frac{1}{\sqrt{2}}(u_+v_- - u_-v_+)
\ee
where $u_\pm$ and $v_\pm$ are eigenvectors of the spin operators in the $z$-direction corresponding to the eigenvalues $\pm1$, and we choose $f_1(z_1)$ and $f_2(z_2)$ to be Gaussian.

We now use our assumption that the fields are strong enough to surpass the kinetic part of the Hamiltonian and write the Pauli equation in the form
\be
i \hbar \frac{\partial \psi}{\partial t}=\left[
 \mu B_1(z_1) \sigma _z\otimes I +\mu B_2(z_2)I\otimes \sigma _z
\right] \psi
\ee
Solving it and taking (\ref{1}) as an initial condition we have
\begin{widetext}
\be \label{2}
\psi(z_1,z_2,t)=\frac{1}{\sqrt{2}}f_1(z_1)f_2(z_2)
\left[
e^{-i\frac{ \mu}{\hbar}(B_0(1-sK)+Bz_1-sKBz_2)t}u_+v_-
-e^{-i\frac{ \mu}{\hbar}(B_0(-1+sK)-Bz_1-sKBz_2)t}u_-v_+
\right]
\ee
It is convenient to Fourier analyze functions $f_1(z_1)$ and $f_2(z_2)$
\be
f_1(z_1)&=&\frac{1}{\sqrt{2\pi}}\intinf dk_1 g_1(k_1)e^{ik_1z_1}
\\
f_2(z_2)&=&\frac{1}{\sqrt{2\pi}}\intinf dk_2 g_2(k_2)e^{ik_2z_2}
\ee
(\ref{2}) becomes then
\be \nonumber
\psi(z_1,z_2,t)
&=&
\frac{1}{2\sqrt{2}\pi}
\intinf dk_1 \intinf dk_2 g_2(k_2)e^{ik_2z_2} g_1(k_1)e^{ik_1z_1}
\nonumber\\
&\times&
\left(
e^{-i\frac{ \mu}{\hbar}(B_0(1-sK)+Bz_1-sKBz_2)t}u_+v_-
-e^{-i\frac{ \mu}{\hbar}(B_0(-1+sK)-Bz_1-sKBz_2)t}u_-v_+
\right)
\ee
\end{widetext}
After leaving the field the system evolves according to the free Schr\"{o}dinger equation:
\be
i\hbar\frac{\partial \Psi}{\partial t}=-\frac{\hbar^2}{2m} \left( \frac{\partial ^2}{\partial z_1^2} + \frac{\partial ^2}{\partial z_2^2} \right) \Psi
\ee
Its general solution is of the form
\be \label{3.5}
\Psi(z_1,z_2,t)= \sum _{a,b=\pm}\intinf dk_1 \intinf dk_2 
 \Psi_{k_1,k_2}^{ab}(z_1,z_2)e^{-\frac{i}{\hbar}E_{k_1,k_2}^{ab}t}u_av_b
\ee
where
\be \label{4}
\left( \frac{\partial ^2}{\partial z_1^2} + \frac{\partial ^2}{\partial z_2^2} \right) \Psi_{k_1,k_2}^{ab}(z_1,z_2)+\frac{2mE_{k_1,k_2}^{ab}}{\hbar^2}\Psi_{k_1,k_2}^{ab}(z_1,z_2)=0
\ee
Now (\ref{3.5}) becomes the initial condition leading to
\be
\psi(z_1,z_2,T)=\Psi(z_1,z_2,0)
\ee
Comparing we get
\be \label{5.1}
\Psi_{k_1,k_2}^{++}(z_1,z_2)&=&\Psi_{k_1,k_2}^{--}(z_1,z_2)=0
\\
\Psi_{k_1,k_2}^{+-}(z_1,z_2)&=&\frac{1}{2\sqrt{2}\pi}
g_2(k_2)e^{ik_2z_2} g_1(k_1)e^{ik_1z_1}
e^{-i\frac{\mu T}{\hbar}(B_0(1-sK)+Bz_1-sKBz_2)}
\label{5.2}\\
\Psi_{k_1,k_2}^{-+}(z_1,z_2)&=&-\frac{1}{2\sqrt{2}\pi}
g_2(k_2)e^{ik_2z_2} g_1(k_1)e^{ik_1z_1}
e^{-i\frac{\mu T}{\hbar}(B_0(-1+sK)-Bz_1-sKBz_2)}
\label{5.3}
\ee
Using (\ref{4}) we calculate:
\be \label{6.1}
E_{k_1,k_2}^{+-}=\frac{\hbar^2}{2m}\left[\left(k_1-\frac{B\mu T}{\hbar}\right)^2
+\left(k_2+sK\frac{B\mu T}{\hbar}\right)^2\right]
\\
E_{k_1,k_2}^{-+}=\frac{\hbar^2}{2m}\left[
\left(k_1+\frac{B\mu T}{\hbar}\right)^2
+
\left(k_2-sK\frac{B\mu T}{\hbar}\right)^2\right]
\label{6.2}
\ee
Substituting explicit expressions for $g_1$ and $g_2$
\be
g_1(k_1)=\left( \frac {2 \sigma_0^2}{\pi} \right)^{\frac{1}{4}}e^{-k_1^2\sigma_0^2} \qquad
g_2(k_2)=\left( \frac {2 \sigma_0^2}{\pi} \right)^{\frac{1}{4}}e^{-k_2^2\sigma_0^2}
\ee
(\ref{5.1})--(\ref{5.3}) and (\ref{6.1})--(\ref{6.2}) to (\ref{3.5}) we have
\be 
\Psi(z_1,z_2,t)
&=&
\frac{\sigma_0}{2\pi \sqrt{\pi}}e^{-i\frac{\mu T}{\hbar}(B_0(1-sK)+Bz_1-sKBz_2)}
\nonumber\\
&\pp=&\times
\intinf dk_1 e^{-k_1^2\sigma_0^2}e^{ik_1z_1}e^{-i\frac{\hbar}{2m}\left(k_1-\frac{B\mu T}{\hbar}\right)^2t}
\intinf dk_2 e^{-k_2^2\sigma_0^2}e^{ik_2z_2}e^{-i\frac{\hbar}{2m}\left(k_2+sK\frac{B\mu T}{\hbar}\right)^2t} u_+v_-\nonumber\\
&-&
\frac{\sigma_0}{2\pi \sqrt{\pi}}e^{-i\frac{\mu T}{\hbar}(B_0(-1+sK)-Bz_1+sKBz_2)}
\nonumber\\
&\pp=&\times
\intinf dk_1 e^{-k_1^2\sigma_0^2}e^{ik_1z_1}e^{-i\frac{\hbar}{2m}\left(k_1+\frac{B\mu T}{\hbar}\right)^2t}
\intinf dk_2 e^{-k_2^2\sigma_0^2}e^{ik_2z_2}e^{-i\frac{\hbar}{2m}\left(k_2-sK\frac{B\mu T}{\hbar}\right)^2t} u_-v_+
\ee
Integrating and rearranging we get
\be \nonumber
\Psi(z_1,z_2,t)=\frac{1}{2\sigma_0\sqrt{\pi \ve}}e^{-i\tan^{-1} \left(\frac{\hbar t}{2\sigma_0^2m} \right) }
e^{i\left( -\frac{(1+s^2K^2)B^2 \mu^2 T^2  t}{2\hbar m \ve} +\frac{\hbar t}{8m \sigma_0^4 \ve}(z_1^2+z_2^2)\right)}
e^{-\frac{1}{4\sigma_0^2 \ve}(z_1^2+z_2^2) -\frac{(1+s^2K^2)B^2 \mu^2 T^2  t^2}{4m^2 \sigma_0^2 \ve}}
\times
\\ \nonumber
\times \left(
e^{-i\frac{\mu T}{\hbar}(B_0(1-sK)+Bz_1-sKBz_2)}e^{-\frac{(z_1-sKz_2)B \mu T  t}{2m\sigma_0^2 \ve}}e^{i\frac{\hbar (z_1-sKz_2)B \mu Tt^2}{4m^2 \sigma_0^4 \ve}}u_+v_- + \right.
\\
\left.
-e^{-i\frac{\mu T}{\hbar}(B_0(-1+sK)-Bz_1+sKBz_2)}e^{\frac{(z_1-sKz_2)B \mu T  t}{2m\sigma_0^2 \ve}}e^{-i\frac{\hbar (z_1-sKz_2)B \mu Tt^2}{4m^2 \sigma_0^4 \ve}}u_-v_+
\right)
\ee
where
$
\ve=1+\frac{\hbar^2 t^2}{4\sigma_0^4m^2}.
$
From the conservation law
\be
\frac{\partial \rho}{\partial t} +\frac{\partial j_1}{\partial z_1}+\frac{\partial j_2}{\partial z_2}=0
\ee
we get currents defined by
\be
j_a=\frac{\hbar}{2mi}\left[ \Psi^\dag \frac{\partial}{\partial z_a} \Psi -\left(\frac{\partial}{\partial z_a}\Psi^\dag \right)  \Psi \right],\quad a=1,2
\ee
Its easy to find that
\be
\rho
&=&\frac{1}{2\sigma_0^2\pi \ve}
e^{-\frac{1}{2\sigma_0^2 \ve}(z_1^2+z_2^2)-\frac{(1+s^2K^2)B^2 \mu^2 T^2  t^2}{2m^2 \sigma_0^2 \ve}}
\cosh\left( \frac{(z_1-sKz_2)B \mu T  t}{m\sigma_0^2 \ve} \right)
\\
j_1 
&=&\frac{1}{2\sigma_0^2\pi \ve}
e^{-\frac{1}{2\sigma_0^2 \ve}(z_1^2+z_2^2)-\frac{(1+s^2K^2)B^2 \mu^2 T^2  t^2}{2m^2 \sigma_0^2 \ve}}\frac{\hbar}{m}\left[ \frac{B \mu T }{\hbar \ve}\sinh \left(\frac{(z_1-sKz_2)B \mu T  t}{m\sigma_0^2 \ve} \right) +\frac{\hbar t z_1}{4m\sigma_0^4 \ve}\cosh \left(\frac{(z_1-sKz_2)B \mu T  t}{m\sigma_0^2 \ve} \right) \right]
\\
j_2
&=&\frac{1}{2\sigma_0^2\pi \ve}
e^{-\frac{1}{2\sigma_0^2 \ve}(z_1^2+z_2^2)-\frac{(1+s^2K^2)B^2 \mu^2 T^2  t^2}{2m^2 \sigma_0^2 \ve}}\frac{\hbar}{m}\left[ -\frac{ sKB \mu T }{\hbar \ve}\sinh \left(\frac{(z_1-sKz_2)B \mu T  t}{m\sigma_0^2 \ve} \right) +\frac{\hbar t z_2}{4m\sigma_0^4 \ve}\cosh \left(\frac{(z_1-sKz_2)B \mu T  t}{m\sigma_0^2 \ve} \right) \right]\nonumber\\
\ee
Introducing velocity as a hydrodynamical variable from the formula
$
j_a=\rho v_a
$
we derive velocities (in $z$-direction) of each particle
\be \label{7}
v_1
&=&\frac{\hbar^2 t z_1}{4 m^2 \sigma_0^4 \ve} +\frac{ B\mu T }{m \ve}
\tanh \left(\frac{(z_1-sKz_2)B \mu T  t}{m\sigma_0^2 \ve} \right)
\\ \label{8}
v_2 
&=&
\frac{\hbar^2 t z_2}{4 m^2 \sigma_0^4 \ve} -\frac{ sKB\mu T }{m \ve}\tanh \left(\frac{(z_1-sKz_2)B \mu T  t}{m\sigma_0^2 \ve} \right)
\ee
In the causal interpretation the particles center of mass has a well-defined translational motion implied by its initial position and velocity. In the case of homogeneous fields only first parts of velocities survive and particles trajectory is hyperbola which results from the natural spreading of the wave packet. For large $B$ and small $t$ the second part dominates and can move the particle to either positive or negative side of the $xy$ plane. When $t\rightarrow \infty$ the second part approaches $0$ and the motion is governed by the first part, but since the sign of velocity is then the same as the sign of the $z$ component the direction of translational motion will not change.

Now we will show how this result can be used to send information.

\section{The Protocol}

1. Alice, Bob or a third party (it is not significant to the protocol) generates $N$ pairs in the singlet state (\ref{1}).

\

2. Bob generates two random numbers $s=\pm1$ and $\delta=\left\{0,\frac{\pi}{2}\right\}$. And aligns his device at an angle $\delta$ to Alice's.

\

3. Let the size of the entrance slit of each of the devices be $d$. Then, basing on the points on the screen where his particles have landed, Bob can calculate whether the position of his particle at the entrance was too close to the center of the device; the criterion is the inequality 
\be
|z_{20}|<\frac{d}{2K}
\ee
If the particle was too close Bob proclaims this result and this instance is disregarded.

\

4. As we have shown above the outcome of the experiment (which side of the $xy$ plane each particle gets to) depends only on the second part of the velocity equation (assuming $B$ is large enough) and, of course, initial position. Thus we neglect first parts in (\ref{7}) and (\ref{8}):
\be
v_1 &=&\frac{B \mu T }{m \ve}\tanh \left(\frac{(z_1-sKz_2)B \mu T  t}{m\sigma_0^2 \ve} \right)
\\
v_2 &=&-\frac{ sKB\mu T }{m \ve}\tanh \left(\frac{(z_1-sKz_2)B \mu T  t}{m\sigma_0^2 \ve} \right)
\ee
It is easy to notice that
\be
W_A &=&
{\rm sgn}(v_1)=-{\rm sgn}(z_{20})s
 \\
W_B &=&{\rm sgn}(v_2)={\rm sgn}(z_{20})
\ee
We will consider $W_A$ and $W_B$ to be outcomes of experiments form Alice's and Bob's points of view. Figures~1 and 2 show the trajectories of pairs of the particles for different values of $s$ (computed numerically). Initial positions have been taken the same in both figures.

\

5. Bob proclaims $\delta$ for every instance of experiment.

\

6. Bob randomly chooses half of the instances of the experiment, and proclaims $W_B$ and $s$ 
($\delta$ is already publicly known). Alice proclaims $W_A$ corresponding to each of proclaimed $W_B$. If in any case for $\delta=0$  $W_B\neq-W_As$ then the procedure is aborted. If not, for both values of $s$ separately, the Bell's expectation value is checked and if it differs from $2\sqrt{2}$ too much the procedure is also aborted.

\

7. We take values $W_A$, for $\delta=0$, to be the bits of the key. Alice knows it since its implied by the side of the screen her particle landed on. Bob knows $s$ so he too can find the bit of the key, as $W_B=-W_As$. Since $s$ is a random number chosen by Bob, not correlated in any way with the state of the system before the particles entered Alice's and Bob's labs, no knowledge about system prior to experiment can lead to the discovery of the key.

\begin{figure}[p] \label{f+}
\includegraphics{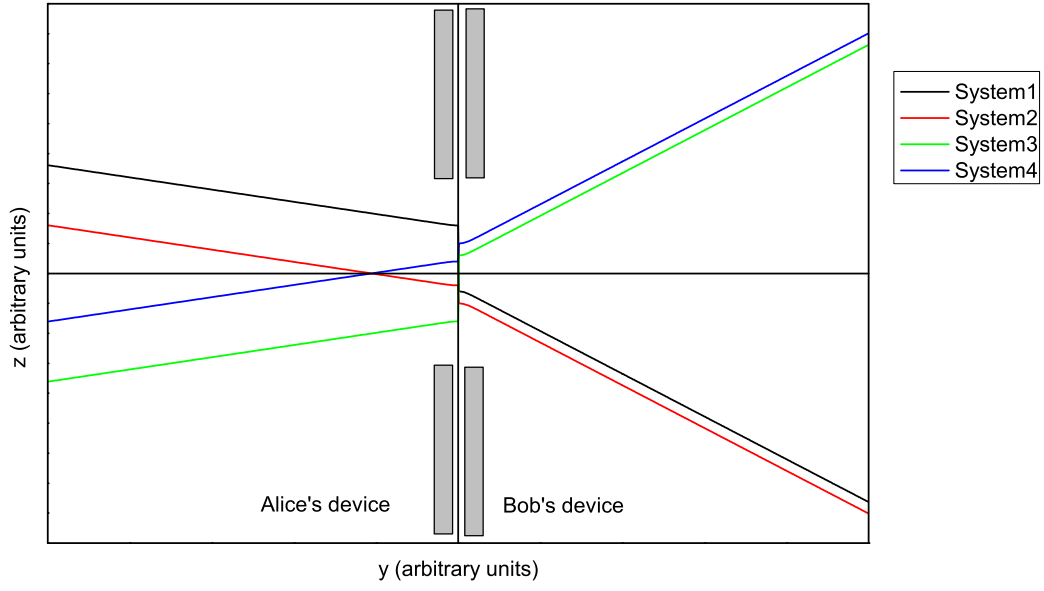}
\caption{Trajectories of particles for $s=+1$. We see that the particles move in the opposite directions along $z$ axis.}
\end{figure}

\begin{figure}[p] \label{f}
\includegraphics{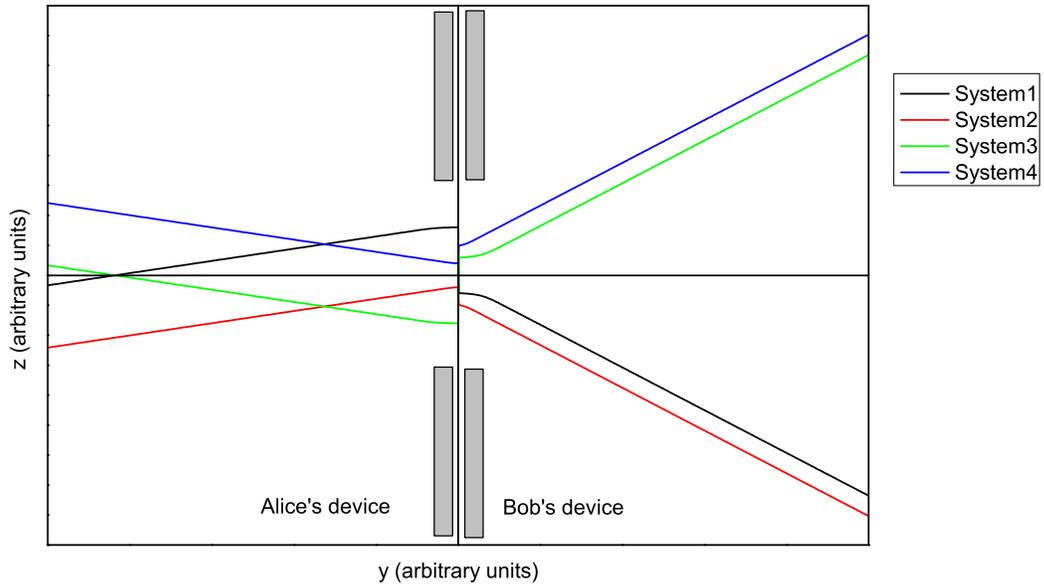}
\caption{Trajectories of particles for $s=-1$. Here Bob's and Alice's particles move in the same direction along $z$ axis. Comparing it with Fig. ~1 we see that changing the value of $s$ has no effect on the outcome of the experiment at Bob's (though the trajectory of his particle is slightly affected) and it flips the outcome of the experiment at Alice's.}
\end{figure}


\begin{thebibliography}{99}
\bibitem{2-time}In \cite{Feligioni} the authors argue that two-time correlation functions computed within the de Broglie-Bohm interpretation will not give the same values as those based on standard quantum mechanics. It can be shown that the result follows from a nonstandard definition of Schr\"odinger-picture correlation functions assumed in 
\cite{Feligioni}. There will be no difference between the two interpretations if one formulates two-time correlation functions in a way introduced in \cite{CD}.
\end{thebibliography}
\end{document}